\documentstyle[12pt,epsf,epsfig,wrapfig]{article}
\begin{document}
\begin{center}
{\large \bf
Spin Effects in Diffractive Deep Inelastic Scattering
\\ }
\vspace{5mm}
S.V.Goloskokov\\
\vspace{5mm}
{\small\it
Bogoliubov Laboratory of Theoretical Physics, Joint Institute for
Nuclear Research, Dubna 141980, Moscow region, Russia
\\ }
\end{center}

\begin{center}
ABSTRACT

\vspace{5mm}
\begin{minipage}{130 mm}
\small
We discuss the contribution of diffractive $Q \bar Q$ production to
the longitudinal double-spin asymmetry in polarized deep--inelastic
$lp$ scattering. We show the strong dependence of the $A_{ll}$
asymmetry on the pomeron spin structure.
\end{minipage}
\end{center}

The study of diffractive events with a large rapidity gap in deep
inelastic lepton--proton scattering at HERA Ref. [1] has given
excellent tools to test the structure of the pomeron and its
couplings. The future polarized diffractive experiments at HERA, HERA
-$\vec N$ and RHIC Ref. [2] might give the possibility to investigate
the spin structure of the pomeron. Then, the question how large the
spin--flip component of the pomeron should be very important.

The pomeron contribution to the hadron high energy amplitude can be
written as a product of two pomeron vertices
$V_{\mu}^{hhI\hspace{-1.1mm}P}$ multiplied by some function
$I\hspace{-1.6mm}P$ of the pomeron. As a result,  the quark-proton
high-energy amplitude looks like
$$
T(s,t)=i I\hspace{-1.6mm}P(s,t) V_{hhI\hspace{-1.1mm}P}^{\mu}(t)
\otimes
V^{hhI\hspace{-1.1mm}P}_{\mu}(t).
$$
The contributions where the pomeron couples to a single quark lead
to a simple matrix structure of pomeron vertex
\begin{equation}
V^{\mu}_{hh I\hspace{-1.1mm}P} =\beta_{hh I\hspace{-1.1mm}P}
\gamma^{\mu}.
\label{pmu}
\end{equation}
 This standard coupling leads to spin-flip effects decreasing
 with energy like $1/s$.

The large-distance loop contributions  complicate the spin
structure of the pomeron coupling. These effects are determined by
the hadron wave function for the pomeron-hadron couplings or by the
gluon-loop corrections for the quark-pomeron case.

The model calculations Ref. [3] give the
following form of the pomeron--proton
vertex
$$
V_{ppI\hspace{-1.1mm}P}^{\mu}(p,r)=m p^{\mu} A(r)+ \gamma^{\mu} B(r),
$$
where $m$ is the proton mass. The ratio of amplitudes $m^2 |A|/|B|$
has been found of about 0.2 for $|t| \sim 1GeV^2$.
The predicted single and double transverse spin asymmetries (Ref.[4])
are about $10 - 15\%$ and have a weak energy dependence. They can
be studied in future experiments at RHIC.

The spin structure of the quark-pomeron coupling
$V_{qqI\hspace{-1.1mm}P}^{\mu}$ has been studied in Ref. [5]. It has
been shown that in addition to the standard pomeron vertex
(\ref{pmu})  the gluon-loop contributions should be important which
have been calculated perturbatively. As a result, the following form
of quark-pomeron vertex has found in Ref. [5]:
\begin{equation}
V_{qqI\hspace{-1.1mm}P}^{\mu}(k,r)=\gamma^{\mu} u_0+2 M_Q k^{\mu}
u_1+
2 k^{\mu}
/ \hspace{-2.3mm} k u_2 + i u_3 \epsilon^{\mu\alpha\beta\rho}
k_\alpha r_\beta \gamma_\rho \gamma_5+i M_Q u_4
\sigma^{\mu\alpha} r_\alpha,    \label{ver}
\end{equation}
where $k$ is the quark momentum, $r$ is the momentum transfer and
$M_Q$
is the quark mass.  The functions
$u_1(r) - u_4(r)$ are proportional to $\alpha_s$ and lead to
spin-flip at the quark-pomeron coupling. These
functions can reach $30 \div 40 \%$ of the standard pomeron term
$u_0(r)$ for $|r^2| \simeq {\rm few}~ GeV^2$ (Ref.[6]).

The new form of the quark--pomeron coupling (\ref{ver}) should modify
various  spin
asymmetries in high--energy diffractive reactions (Ref. [7,8]).
In this report we shall discuss spin asymmetries
in  diffractive $ep$ reactions, which may be analyzed
in the future  polarized HERA experiments to test the spin structure
of the pomeron.

\begin{wrapfigure}{r}{7.5cm}
\epsfig{figure=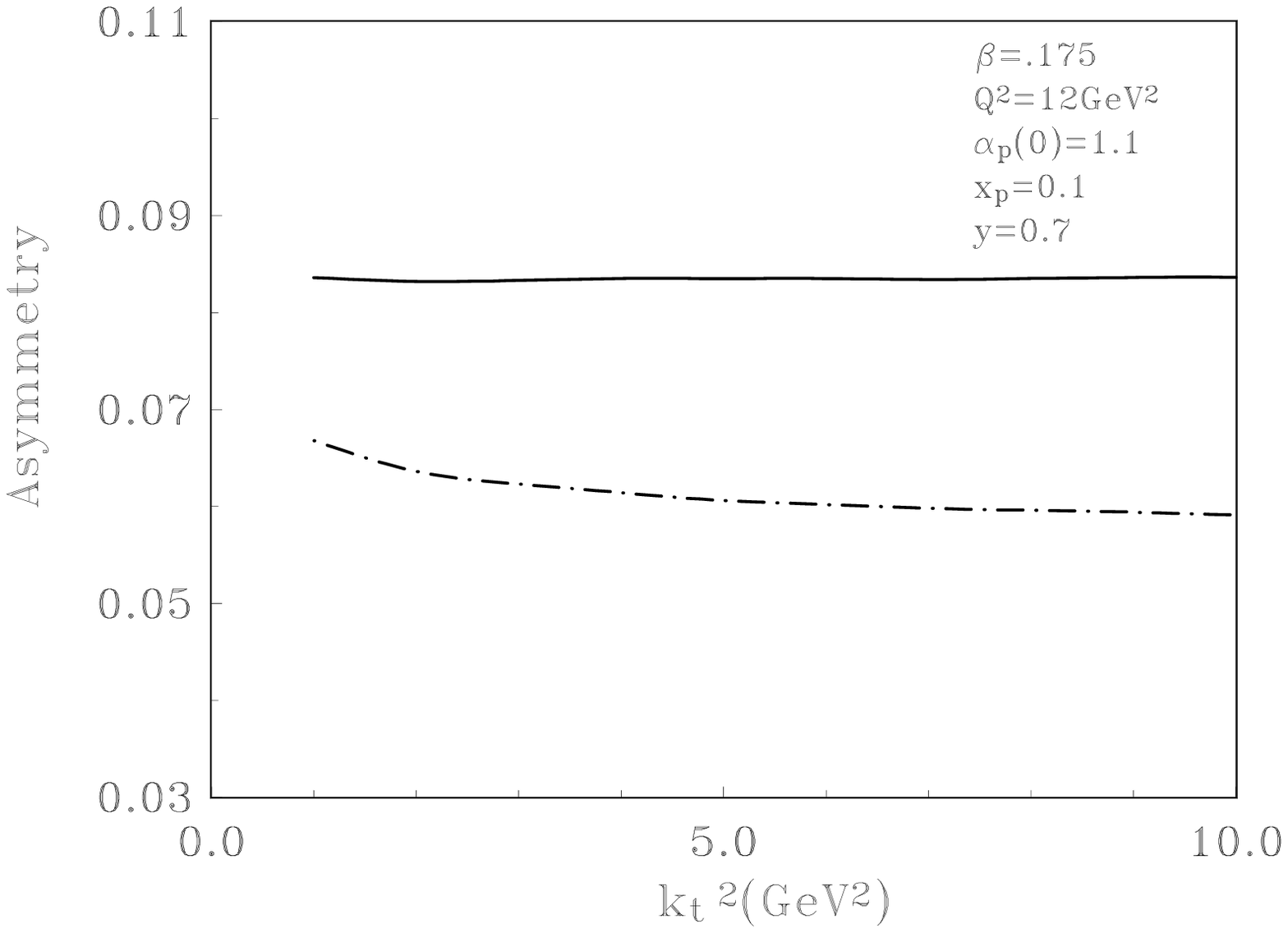,width=7.8cm}
{\small Figure 1: $k^2_{\perp}$-- dependence of $A_{ll}$ asymmetry.
Solid line -for the standard vertex;
dot-dashed line -for the spin-dependent quark-pomeron vertex.}
\end{wrapfigure}

Let us study the effects of spin--dependent quark--pomeron
coupling  (Ref. [9]) in diffractive deep inelastic scattering.
The longitudinal double spin $A_{ll}$ asymmetry  is
determined by the relation
$A_{ll}=
(\sigma(^{\rightarrow} _{\Leftarrow})-\sigma(^{\rightarrow}
_{\Rightarrow}))/
(\sigma(^{\rightarrow} _{\Rightarrow})+\sigma(^{\rightarrow}
_{\Leftarrow}))$
where $\sigma(^{\rightarrow} _{\Rightarrow})$ and
$\sigma(^{\rightarrow}
_{\Leftarrow})$ are the cross sections with parallel and antiparallel
longitudinal polarization of lepton and proton.	 We calculate here
the cross section
integrated over the momentum transfer at the pomeron-proton vertex.

The results of calculation for
the $A_{ll}$ asymmetry of the light quark production in diffractive
deep
inelastic scattering for $\beta=0.175, Q^2=12(GeV)^2, x_p=0.1, y=0.7$
and
the pomeron intercept $\alpha_{P}(0)= 1.1$  are shown in Fig. 1 for
the
standard and spin-dependent pomeron couplings.

It is found that the $A_{ll}$ spin asymmetry of
diffractive
$Q\bar Q$ production for the standard quark--pomeron vertex is very
simple
in form (Ref. [9])
$$
A_{ll}=\frac{y x_p (2-y)}{2-2y+y^2}.$$
 For the
spin--dependent pomeron coupling the asymmetry is dependent on
$k^2_{\perp}$ of jets and smaller than for the
standard pomeron vertex. Thus, one can use the $A_{ll}$ asymmetry to
test
the quark-pomeron coupling structure.

From the difference of the polarized cross sections  the diffractive
contribution to the spin--dependent structure function $g_1$ can be
defined as
$$
\frac{d^4\sigma(^{\rightarrow} _{\Leftarrow})}{dx dy dx_p dt}
-\frac{d^4\sigma(^{\rightarrow} _{\Rightarrow})}{dx dy dx_p dt}
=\frac{8 \pi \alpha^2}{Q^2} (2-y) g^D_1(x,Q^2,x_p,t).
$$

The diffractive contribution to $g_1$ can be found
from the integrated $g^D_1(x,Q^2,x_p,t)$ structure function

\begin{wrapfigure}{r}{7.5cm}
\epsfig{figure=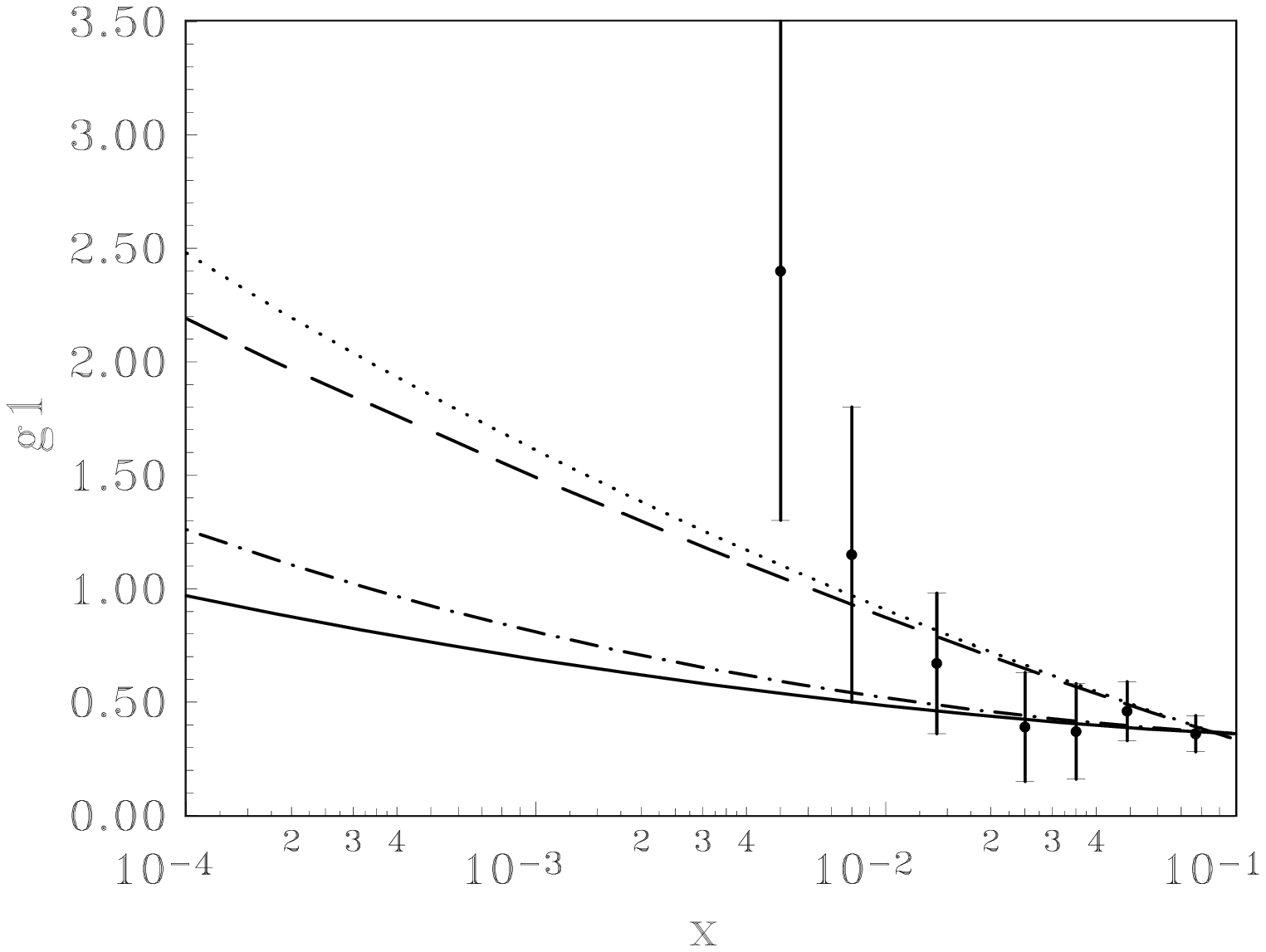,width=7.8cm}
{\small Figure 2: 
The sum of the constant and diffractive $g^D_1$ contributions to
$g_1$:
solid line -- for $\alpha_{P}(0)=1.1$, dot-dashed line
 -- for $\alpha_{P}(0)=1.15$.
The sum of $g^D_1$ and the pomeron contribution (Ref. [11])
to $g_1$:
dashed line -- for $\alpha_{P}(0)=1.1$, dotted line
 -- for $\alpha_{P}(0)=1.15$.
Data are from Ref. [12]}.
\end{wrapfigure}

$$g^D_1(x,Q^2)=\int dx_p dt g^D_1(x,Q^2,x_p,t).$$
The obtained low-$x$ behaviour of $ g_1$ has a
singular
form like $1/(x^{0.3} \ln^2(x))$  (Ref. [10]) which is compatible
with the SMC data for $g^p_1$ at $Q^2=10(GeV)^2$ (Fig.2).

We can conclude that the study of the longitudinal double
spin asymmetry in diffractive deep inelastic
scattering can give an important information about the  spin
structure of the pomeron coupling. For testing the pomeron--proton
vertex the transverse polarization of the proton  should be
more relevant. The future spin facilities at HERA is one of the best
places to
study properties of the pomeron and to test the size of its
spin--flip component.

\vspace{0.2cm}
\vfill
{\small\begin{description}
\item{[1]}
H1 Collaboration,  T.Ahmed et al, Phys.Lett.  {\bf B348} (1995)
681;\\
ZEUS Collaboration,  M.Derrick et al.  Z.Phys. {\bf C68} (1995) 569.
\item{[2]}
 W.-D.Nowak,  11th International Symposium on High Energy Spin
Physics,
Bloomington in 1994, ISBN 1-56396-374-4,
AIP Conference Proceedings {\bf 343} (1995) 412;\\
A.Sch\"afer and J.Feltesse, to be published in Proc of the Workshop
 "Future
Physics at HERA", DESY May 29-31, 1996.
\item{[3]}  S.V.Goloskokov, S.P.Kuleshov, O.V.Selyugin,
           Z.Phys.  {\bf C50} (1991) 455.
\item{[4]}
S.V.Goloskokov, O.V. Selyugin, Yad. Fiz.  {\bf 58} (1995) 1894.
\item{[5]}
S.V. Goloskokov, Phys. Lett.  {\bf B315} (1993) 459.
\item{[6]}
S.V.Goloskokov, O.V. Selyugin, Yad. Fiz.  {\bf 57} (1994) 727.
\item{[7]}
S.V.Goloskokov, Phys.Rev. {\bf D53} (1996) 5995
\item{[8]}
J.Klenner, A.Sch\"afer, W.Greiner, Z.Phys. {\bf A352} (1995) 203.
\item{[9]}
S.V. Goloskokov, E-print hep-ph 9604359
\item{[10]}
S.V. Goloskokov, E-print hep-ph 9604261.
\item{[11]}
S.D.Bass, P.V.Landshoff, Phys. Lett. {\bf B336} (1994) 537.
\item{[12]}
SMC Collaboration, D.Adams et al., Phys. Lett. {\bf B357} (1995) 248.
\end{description}}

\end{document}